# On Using Synthetic Social Media Stimuli in an Emergency Preparedness Functional Exercise

Andrew Hampton[1], Shreyansh Bhatt[2], Alan Smith[2], Jeremy Brunn[2], Hemant Purohit[2], Valerie L. Shalin[1], John M. Flach[1], & Amit P. Sheth[2]

[1] Kno.e.sis, Wright State University, Department of Psychology
[2] Kno.e.sis, Wright State University, Department of Computer Science & Engineering

Corresponding author: Andrew Hampton
hampton.41@wright.edu

**Abstract**
This paper details the creation and use of a massive (32,000+ messages) artificially constructed 'Twitter' microblog stream for a regional emergency preparedness functional exercise. By combining microblog conversion, manual production, and a control set, we created a web-based information stream providing valid, misleading, and irrelevant information to public information officers (PIOs) representing hospitals, fire departments, the local Red Cross, and city and county government officials. PIOs searched, monitored, and (through conventional channels) verified potentially actionable information that could then be redistributed through a personalized screen name. Our case study of a key PIO reveals several capabilities that social media can support, including event detection, the distribution of information between functions within the emergency response community, and the distribution of messages to the public. We suggest that training as well as information filtering tools are necessary to realize the potential of social media in both emergencies and exercises.

## 1. Introduction

There is a growing consensus that social media has the potential to play a role in the management of regional disasters (e.g., a terrorist bombing or a weather event such as a tornado, forest fire, or hurricane) [3, 5, 6, 7, 9, 4, 8, 10, 11, 12, 13, 15, 16, 17]. Social media act as a platform for citizen sensing [18] that can be a source of timely information to emergency command centers about events happening on the ground [13], changing availability and needs for resources [6, 7, 12], and changing public perceptions of events [3]. Social media can also be a means for communicating with the larger public (e.g., to alert the public to evolving risks; to instruct the public about appropriate responses and available resources; to correct rumors; and to solicit information or resources) [3, 5]. Thus, social media based exchange between the informal (citizen-based) and formal (professional) communities may facilitate effective crisis management [7, 17].

Although potential roles for citizens in emergency response appear useful, and despite recent initiatives by the formal response community such as the U.S. Federal Emergency Management Agency [1], the command and control models from the formal response organizations do not easily accommodate the social media that the informal response community has adopted. While a growing research base examined the response and reaction of the public to social media in an emergency response context, limited data address the functionality of social media for the formal emergency response organization. This paper will report on our experiences integrating a social media component within a functional training exercise for a local emergency in a medium-size city in the Midwestern United States. It involved hospitals, county public health departments, city government, fire departments, police departments, and the regional Red Cross. Certainly, any effort to deploy a social media tool in a high consequence setting requires extensive testing, ideally conducted with the fidelity of a training exercise. However, the examination of social media usage in a training exercise has additional advantages. Because an exercise is designed, we had access to ground truth. We were also able to control the message stream, and therefore knew the distribution of its contents and the rate of presentation. Known threats to usability such as data overload and low signal-to-noise ratio, that can be addressed with technology (e.g., Twitris [19]), can therefore be reduced while we examine utility. However, this control required the design of the social media stream synchronized with designed exercise content, and a standalone tool to restrict the distribution of messages to exercise participants.

**2. The Exercise**

According to the Emergency Management Institute's Independent Study Program course on emergency management exercises [2], there are seven distinct types of exercises ranging from seminar to full-scale simulation. We conducted the current work in the context of a 'functional exercise.' This is the second most complex (and realistic) operations-based exercise just short of the full-scale exercise. Functional exercises specifically aim to assess the current response plans and to evaluate regional decision-making processes associated with implementing the plans. Thus, the local (e.g., hospital information centers, fire departments) and region-wide (e.g., Red Cross) command centers actively participated. However, the actual deployment of 'boots on the ground' (e.g., doctors, nurses, fire units, etc.) was simulated. In the following subsections we describe the exercise design team and the details of the scenario.

*2.1 Exercise Design Team and Roles*

The local hospital association coordinated the exercise planning, in conjunction with regional partners such as the local Red Cross and fire department representatives, as part of emergency planning and training requirements for the region. The specific targets for training and evaluation included medical surge, mass casualty, emergency operations and coordination, responder safety and health, WMD/HAZMAT response and decontamination, and coordination of public information and warning. Our research team was included in the exercise planning team, where we functioned as participant observers. Certainly, our participation in the general planning was mandatory in order to develop a properly synchronized social media stream. This participation also provided unique insight into the pedagogical goals of the exercise and the internal dynamics of the emergency planning and decision making process. Members of the larger emergency management community greeted members of the team favorably at conferences and meetings upon hearing about the integration of a social media component (D. Gerstner, personal communication, 9 April 2014). Planners assisted us with the identification of relevant (or purposely misleading) content for social media streaming. However, we frequently contributed general (that is, not social media dependent) items to the Master Scenario Event List (MSEL) alongside the conventional team as equal partners.

We consulted separately with the social media coordinator for the city at the center of the exercise in order to ascertain specifics of his job description and his methods of interaction. He functioned as a guide for our understanding of the PIO position at the highest levels of functionality and he prevented us from undershooting in terms of complexity and capability. This PIO is among the most engaged with social media (as measured by either followers or interaction, relative to other PIOs in the exercise). He did not have any information on the specifics of the exercise and was subject to detailed analysis during the exercise.

*2.2 Scenario*

The scenario included two separate radiological attacks by a terrorist group. The first attack involved low level radiation distributed through the ventilation system at a large concert for children on Friday evening (the night before the exercise). The second attack was a radiological explosion (dirty bomb) at a large international convention on Saturday morning. This attack involved hundreds of injuries/fatalities in the explosion as well as widespread exposure to potentially lethal levels of radiation.

Designed to encompass roughly three and a half hours, the scenario began early Saturday morning when radiation detection alarms go off at two different local hospitals (approximately 7:55 a.m.). The people triggering the "portal alarms" had attended the Friday evening concert and were coming to emergency rooms due to illness or injuries that were unrelated to the radiation events. Standard protocol is for the patient who sets off these alarms to be escorted out of the emergency room and brought back into the hospital through an alternative route to avoid potential contamination of the emergency room. This high degree of uncertainty about the source of the radiation tested coordination across hospitals. Shortly thereafter (approx. 8:10 a.m.) the dirty bomb explodes, testing medical surge and mass casualty response.

As the scenario created uncertainty about the sources of radiation and the potential lethality of the radiation by design, a clear potential for confusion and public panic resulted. Thus, the exercise represented an ideal context for using social media to simulate the public confusion in a real disaster.

**3. The Microblog Stream Contents**

We employed a Twitter-like system of microblogs. Each entry followed the typical format of a 140 character microblog, and included an indication of the sender, message content, and hashtags to group messages thematically. Each microblog also had a visibility attribute, set to *high*, *medium* or *low* by the research team. This controlled the proliferation of microblogs throughout the exercise, and reflected either sender influence or context relevance. We designed the software to randomly select usernames from an extensive list of real Twitter users for each artificially generated message. Users could click these names in the interface that would then hyperlink to those real user profiles on Twitter in a new window. For the actual scenario participants, the link directed to their professional profile.

The microblog stream of tweets consisted of three distinct sources:

1) *Background set*: the real-time stream of microblogs unrelated to the event by actual Twitter users in the region (background noise);

2) *Authoritative set*: those entered at the user interface by PIOs and simulation coordinators, and their 'retweets' (forwarded tweets) generated by the system; and

3) *Constructed set*:

3a) *MSEL specific constructions*: those generated by the system as required for the MSEL;

3b) *Generic emergency related constructions*: specified by parameterized microblog templates for categories such as 'Angry Rant'.

Table 1 lists the semantic categories that we used to generate specific microblogs. These categories reflect both content developed in the exercise planning process and general content reflecting the analysis of previous actual emergencies. In particular, we considered two major events of 2013: the Boston Marathon Bombing [3] and the Westgate Mall attack [Card et al., 2013]. Prior studies of these events provide examples of real microblogs on Twitter and categories of content discussed in the aftermath of these events. While both incidents were important in shaping our understanding of message categories, the Boston Marathon bombing provided a more directly compatible dataset, as our simulation scenario also involved the bombing of a metropolitan American city during a large gathering of international participants. We sampled a set of tweets from the Twitter dataset that we had collected during the Boston Marathon bombing via the Twitris system (http://twitris.knoesis.org). This constantly collected the filtered stream of English language tweets from the Twitter Streaming API for event-related keywords and hashtags (e.g., "#bostonbomb", "boston bomb", "boston marathon"). The sample included all tweets originating from the state of Massachusetts in approximately 2.5 hours after the disaster took place, and resulted in 2567 messages.

Table 1

*Semantic categories for scenario relevant microblogs. Notation reflects the distribution in our corpus: T = Total Tweets, H=High visibility, M=Medium visibility, L=Low visibility where T=H+M+L.*

| | | |
|---|---|---|
| Angry rant<br>T=149, L=85, M=64 | Distant Observer<br>T= 426, H=426 | Panic over exposure<br>T=18, L=18 |
| Appeal for help<br>T=71, M=71 | Fear for children * 113<br>T=113, H=62, M=32, L=19 | Parent who set off alarm<br>T=131, L=131 |
| Breaking news – explosion<br>T=585, H=426, M=73, L=86 | General discussion<br>T=492, H=281, M=141, L=70 | Prayer<br>T=93, L=93 |
| Breaking news - status update<br>T=72, M=72 | Ham operators with nowhere to go<br>T=88, H=40, M=48 | Public reaction<br>T=31, L=31 |
| Call for calm/patience<br>T=63, H=43, M=20 | Informational<br>T=190, H=190 | Revision of all-clear<br>T=120, H=60, M=60 |
| Call for help<br>T=40, M=40 | Injured<br>T=114, M=71, L=43 | RRR TWEET * 182<br>T=180, L=182 |
| Caution and Advice<br>T=78, M=60, L=18 | JohnQPublic<br>T=140, H=140 | Rumor/ False information<br>T=130, M=130 |
| Confusing public reaction on reports T=50, L=50 | Media help resources<br>T=25, M=25 | Status update - radiation<br>T=905, H=905 |
| Confusion of Hara/Hobart<br>T=96, H=64, M=32 | Media report<br>T=175, H=122, M=53 | Sympathy<br>T=40, M=40 |
| Confusion of TC-99(m)<br>T=30, M=30 | Non-immediate witness<br>T=255, H=141, M=71, L=43 | Uninjured present<br>T=43, L=43 |
| Correct information/treatment<br>T=66, M=66 | Non-immediate witness, uninformed<br>T=71, H=141, M=71, L=43 | Uninjured, injured friend<br>T=43, L=43 |
| Corroboration * 213<br>T=213, M=168, L=50 | Observers in ER waiting room<br>T= 514, M=320, L=194 | What do I do?<br>T=183, H=62, M=102, L=18 |
| Criticism<br>T=80, M=80 | Offer to help<br>T=58, M=58 | Where to go?<br>T=144, H=60, M=66, L=18 |
| Dayton Region<br>T=828, M=828 | Official announcement<br>T=240, H=240 | Worried about exposure<br>T=505, H=161, M=245, L=99 |

*3.1 Background Set*

To simulate the background noise (microblogs unrelated to the emergency events) we collected tweets from the Dayton region over a four-day window that included no particularly noteworthy events (i.e., no major elections, sporting events, conferences, etc.). We used the Twitter Streaming API's public "statuses / filter" [20] method to obtain tweets from the specific region of the exercise using a bounding box defined by a pair of latitude and longitude for the two points on the geographical map of the region: south-west and north-east. Visibility for the background set was low. The background set comprised approximately 76% of the microblog stream.

*3.2 Authoritative Set*

We also had the capability to generate additional microblogs in real time during the exercise, manually specifying the relative visibility of each message as high, medium, or low, corresponding to the visibility levels we assigned to each PIO account based on real-world followership. These messages appeared in the unfiltered social media feed without distinction and were 'retweeted' by ghost accounts, the frequency and duration of which determined by their relative visibility setting, set at the time of creation based on a subjective conjecture about how popular a post of that type would be.

*3.3 Constructed Set: MSEL Specific Constructions*

Scenario specific microblogs included the following examples.

Example 1: patient at *[hospital]* just kicked out after testing positive for radiation… WTF!?!
Example 2: They just kicked some guy out of the ER for radiation poisoning #HulkingOut
Example 3: Do hospitals not treat radiation poisoning? Am I in trouble?

These microblogs simulated witnesses in the emergency room, who sent these when someone who had been at the concert on the previous evening tripped the radiation detection alarms. Because the witnesses are not privy to hospital protocol for rerouting such cases, they infer that a patient is being denied service at that hospital due to radiation. Also included (see Example 2) is a hashtag that ran throughout the simulation at various points. This particular one demonstrates contradictory levity that appears in real tweet streams. These examples deliberately indicate a lack of understanding on the part of the public, the correction of which falls within the bailiwick of public information officers (PIOs). Visibility for these messages depended upon designed ground truth. True events were set to high, and others as required by the exercise for the relevant semantic category of the event (see Table 1).

*3.4 Constructed Set: Generic Emergency-Related Constructions*

In addition to specific microblogs that were tied to specific scenario events, we added microblogs based on observations of Twitter streams taken from actual emergency events. Initially, we attempted to directly transform observed tweets to the exercise scenario by using string-based transformation rules. Obvious examples (e.g., #bostonbombing became #daytonbombing) presented little trouble, but the depth of contextualization made a complete transformation both prohibitively difficult and occasionally erroneous. For example, discussions of the marathon could have related terms like runners and mile markers (and permutations of both). Eventually we concluded that the extensive details of the context precluded this approach.

Instead of this direct transformation, we opted for a thematic translation of the events. Looking at the messages and the type of categories observed in prior studies [3, 4], we created templates of microblogs. Some we made by transforming items within tweets from Boston (see Example 4). For others, we adapted themes to the specific context of our simulation (see Example 5). We then varied the options in parentheses to create a set of functionally similar but identifiably distinct microblogs. In Example 5, the two alternatives in two places (denoted by forward slashes and parentheses, respectively) yielded four possible messages to create a computationally inexpensive diversity.

Example 4: RT @the123abc: It should be noted that if you suffer from #PTSD, limit your exposure to the **(#daytonbomb / #radiation)** coverage. Social media can b overwhelming.
Example 5: **(#bananasplits / #hobartarena) (We enjoyed with kids / kids loved)** the concert at Hobart last night

Visibility depended upon the relevance of the semantic category of microblog content for the MSEL timeline. Because we were also part of the exercise design team and had access to ground truth, we had knowledge of relevance. In Example 4, we set visibility to 'medium' given that it was related to the "caution and advice" category, and inherited from the transformation of a Boston bombing related tweet set.

## 4. Integration of Microblog Contents Sources

We integrated the scenario-related microblogs and background noise to create a corpus of about 32,000 microblogs. We then synchronized the timing of scenario-related microblogs with the MSEL (Master Scenario Event List) for the exercise. During the exercise the microblogs streamed at a rate of approximately two per second.

The scheduling of the frequency and volume of microblog contents was designed to reflect the specified visibility level for individual microblogs. The fact that the number of microblogs from the background set and the authoritative set and subsequent 'retweets' was unknown prior to the start of the simulation presented an integration challenge. To address this challenge, our primary scheduling heuristic made use of the values generated by a real-time trending topics function from Twitris system [Sheth et al., 2014] in the overall data stream at each timestamp 't', throughout the simulation. We considered the count of microblogs mentioning the topmost trending topic (within the current sliding window of past K minutes the timestamp 't') to be the current baseline for high visibility, from which the values for medium and low visibility were proportionally derived. We defined medium visibility as one half of the value of high visibility, and low visibility was one third of value for high visibility. This final count was used as the volume of microblogs scheduled to appear at random intervals throughout the next five minute window, starting from the entry time of the initial template microblogs from the Constructed Set. These fractional values gave us the desired effect of topics mentioned in a high visibility microblog template or PIO user microblog, to appear at or near the top of a list of the top 20 trending topics. Those from a medium visibility microblog template probably (but not necessarily) appeared somewhere within the list, and those from a low visibility microblog template probably not (but still possibly) appearing in the list non-deterministically.

While the approach of using basic proportionality based functions to derive volumes for medium and low visibility (where k=½ and k=⅓, respectively) seemed to adequately meet the visibility goals for our scenario, these functions should be considered parameters which may need to be adjusted for other synthesized streams in future exercises, based on differences in distributions of topics and visibility levels. Additionally, the sliding window size of five minutes should also be considered a parameter which may need to be adjusted with respect to the relative velocity of the stream. In scenarios which involve a much larger number of users, it is anticipated that the exceedingly high velocity may warrant the substitution of sliding window frequency-based trend detection with a more sophisticated trend detection algorithm.

## 5. The PIO Interface

We developed a password protected web-based interface attempting to implement all of the major functionality of Twitter (see Figure 1). This functionality included the ability to compose and send messages (via a pop-up window accessible through the "Tweet" button), monitor trending topics within the feed, identify specific messages related to a topic (either selected from the trending list or manually input via search box), and view an unfiltered stream updating from top to bottom. Scroll bars on the streaming tweets allowed participants to review previous content outside the field of view. Participants could also highlight and copy microblogs in both the filtered stream and the unfiltered stream. Clicking on usernames provided profile information. In addition to these standard capabilities, we included a map function that localized all pre-constructed microblogs within a user-specified topic. Selecting a push-pin on the map revealed the corresponding microblog. Users could also select an area of the map (the circle in Figure 1) and limit newly incoming microblog pushpins to those originating from that geographic designation. PIOs could also elect to disclose the location at which they composed a microblog.

A banner at the top of the screen reminded the users and anyone who may have happened to see the screen that all messages were simply part of a test and therefore were not to be distributed. We attempted to use excessive caution in this regard as, according to Kevin Sur, Emergency Management Instructor for Ohio Emergency Management Agency (personal communication, 23 May 2014), at least once before in an emergency exercise a PIO had confused real with simulated methodologies, to the point of attracting attention from local news sources.

Each PIO received a username and password. Upon reaching the website, they selected their respective "handle" (which we constructed to reflect name and professional affiliation) from a list and then reached the page displayed in Figure 1. Although not announced to the exercise participants, our team had also developed a mobile device application.

## 6. Observations

Twenty-eight PIOs participated in the social media aspect of the exercise representing most of the agencies involved. We were able to obtain a workstation recording for the central PIO of the exercise. He represented the largest participating entity by population (the city) and also had by far the largest following on Twitter and number of tweets produced, suggesting a high level of engagement and expertise. As such, our research team positioned him in the simulation cell (i.e., control center) and we stationed an adjacent observer to provide field notes. We obtained message data from six of the participating PIOs, intervention from other exercise participants, and follow-up debriefing in the form of a questionnaire and training hotwash. Together these reveal the types of interaction that social media affords. A bug in the microblog-sender association function appeared at 8:37, was announced to participants and was repaired by two hours into the simulation. As a result, several hundred pre-constructed microblogs appeared to come from the PIOs, particularly our primary PIO (as he was the most active). We note adjustments to the data analysis where relevant.

We report our observations below, with respect to the interface functions, microblogging across the PIOs and other social media interventions.

*6.1 PIO interface functions*

We reviewed the screen capture for a detailed account of the use of social media. Evidence for engagement occurs in the form of intentional cursor movement, associated with a change in the appearance of the display. This includes selecting trending topics, scrolling the microblog stream to review previous content, and highlighting microblog stream contents. It also includes using the cursor to point at individual microblogs, presumably to keep place while reading. Prior to the appearance of the naming bug and after it was fixed (approximately a total of one hour and forty minutes of usage), the recorded PIO was nearly completely engaged in the social media tool. We identified a rate of approximately 3.4 social media related cursor displacements per minute during the first 38 minutes, and 3.2 social media related cursor displacements per minute during the first 35 minutes following correction of the naming bug. This metric underestimates engagement, as we did not count multiple movements within the same episode separately (e.g., scrolling the microblog stream back and forth several times without stopping). We note the exploitation of several different functions of the PIO interface.

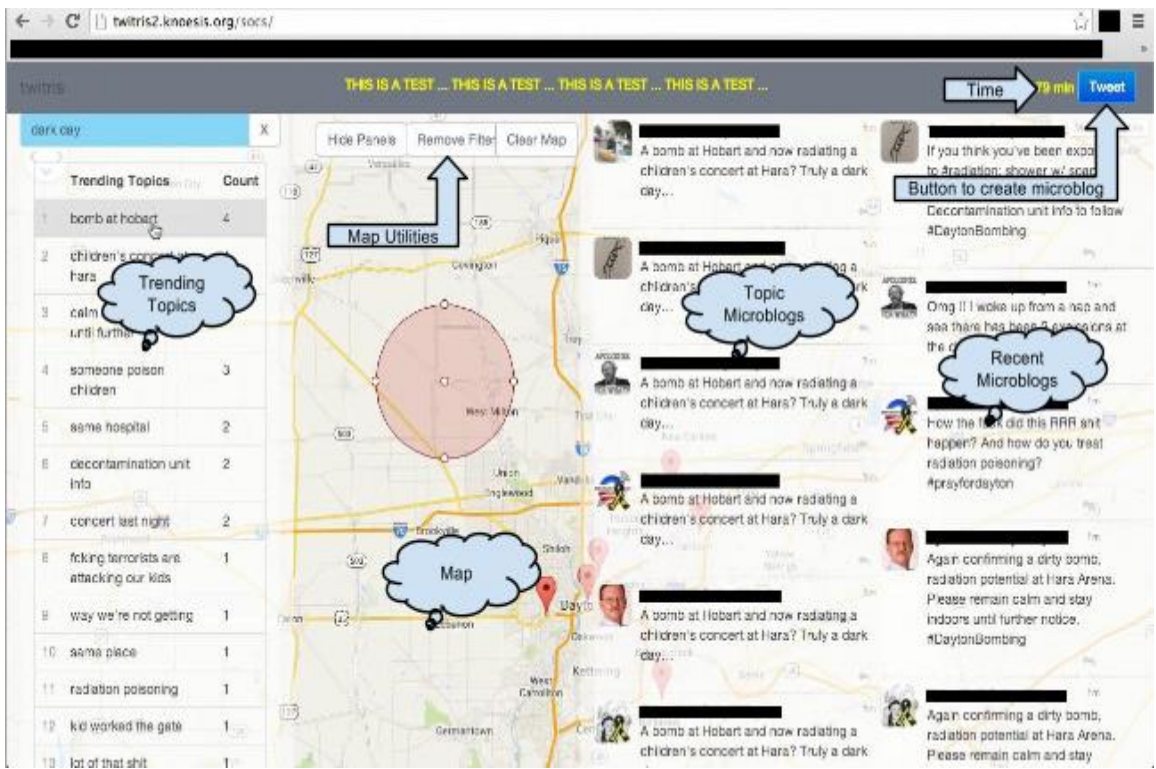

*Figure 1.* PIO Interface. From left to right, Trending Topics: trending topics and phrases; Map Utilities: a map of localized microblogs including a user-designated area of interest; Topic Microblogs: microblogs/tweets under a specified trend; Recent Microblogs: unfiltered stream of all microblogs; Button to create microblog: The button allows users to write microblogs; Time: current duration of the exercise.

*Information monitoring.* The PIO identified microblogs in the unfiltered stream concerning radiation poisoning less than a minute after the first related microblogs appeared. The PIO also identified microblogs in the unfiltered list concerning the dirty bomb, approximately 1'30" after the first related microblog appeared, and almost a minute after the topic first appeared midway in trending topics. Once he detected the event in the unfiltered stream he returned to trending topics and selected a related topic at the top of the list. Both episodes illustrate a reliance on the unfiltered stream for initial detection, and suggest concern for the reliability of the filtering function. Comments in the follow-up survey include his interest in knowing more about how the filtering function works. Nevertheless, the PIO used trending topics to identify the existence of a video of the bombing from the trending topics. He later investigated the hashtag identifier of the potentially responsible group, which did not appear in trending topics. He also relied on trending topics to detect public concern once the name of the radioactive agent was released. Late in the session, once focused on a citizen microblog, the PIO pursued the URL link it contained.

*Information distribution.* The central PIO functioned as a filter on the social media stream, distributing contents via microblogging, e-mail and telephone. We cover his microblogging behavior separately and focus on e-mail and telephone distribution here. The persisting microblogs allowed the PIO to copy microblogs and forward them via e-mail to interested parties including the fire department and the police (see Figure 2). He used this to identify public perception and even provide crime scene photos (see Figure 3). The separate trending topic microblog list allowed the PIO to copy several examples in one step, as opposed to copying individual examples from the more rapidly scrolling unfiltered stream. Nevertheless, no member of the formal response community responded to the PIOs bulletins using e-mail

The PIO also made numerous calls related to the Twitter stream to the fire command center. The function of these calls was typically to check the validity of information gleaned from the social media feed, or at least to make the fire department aware of trending topics he felt merited attention or follow-up. An advantage of the telephone medium is a confirmation of receipt, absent in the one-way e-mail attempts noted above.

*Information source.* The tool provides two functions for examining the source of information: the map and the link to user profiles. Early on, before the dirty bomb and with a preponderance of noise, the PIO manipulated the map field of view, and frequently cleared the push pins. He also used the user profiles to examine the background of the sources before sending microblogs and evidence via e-mail.

Several hours into the exercise, a high visibility account sent out a message warning of a bomb threat at a local community college.

Example 6: "Man with ticking backpack @*[local]*CC. Police evacuate @RedCross shelter. #daytonbomb".

This was not part of the exercise, and subsequent attempts at investigation have failed to identify the author. Upon seeing this in the unfiltered tweet set, the city's PIO immediately called police and fire contacts as well as a simulated FBI agent (who received several calls pursuing the threat) to validate the information. After finishing these calls and coordinating with the account owner, the PIO issued a social media statement that the area was clear and the threat had only been a rumor. This entire episode took fifteen minutes to resolve. While the detection of rumor is encouraging, were this an actual emergency, rumor detection may not resolve the resulting public reaction to an alarming message from an ostensibly reputable source.

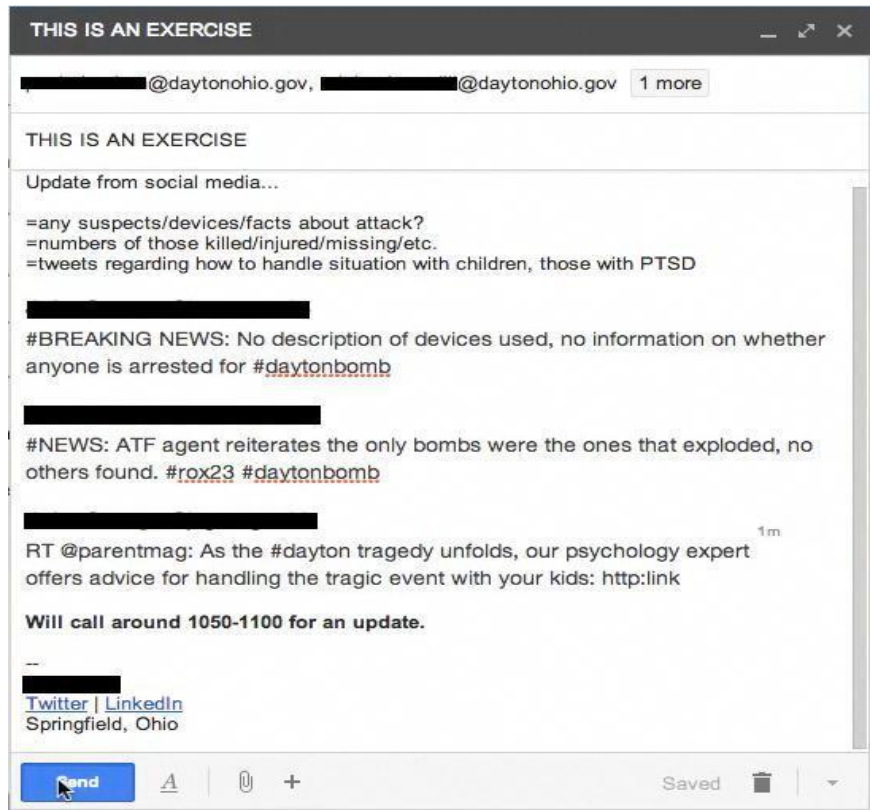

*Figures 2.* Email written by the central PIO to fire department officials.

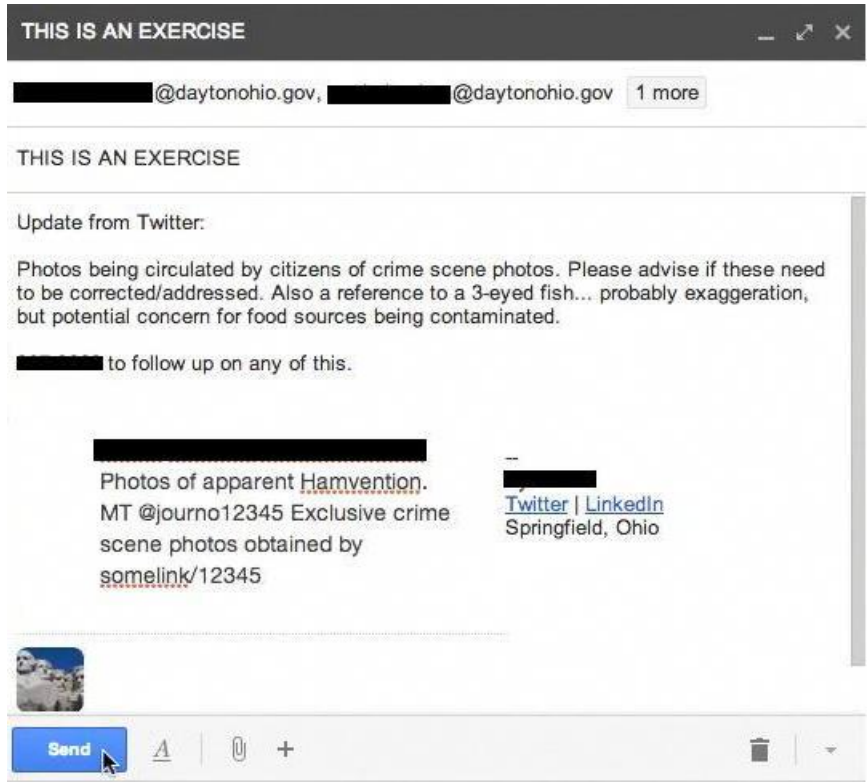

*Figures 3*. Email written by the central PIO to fire department officials.

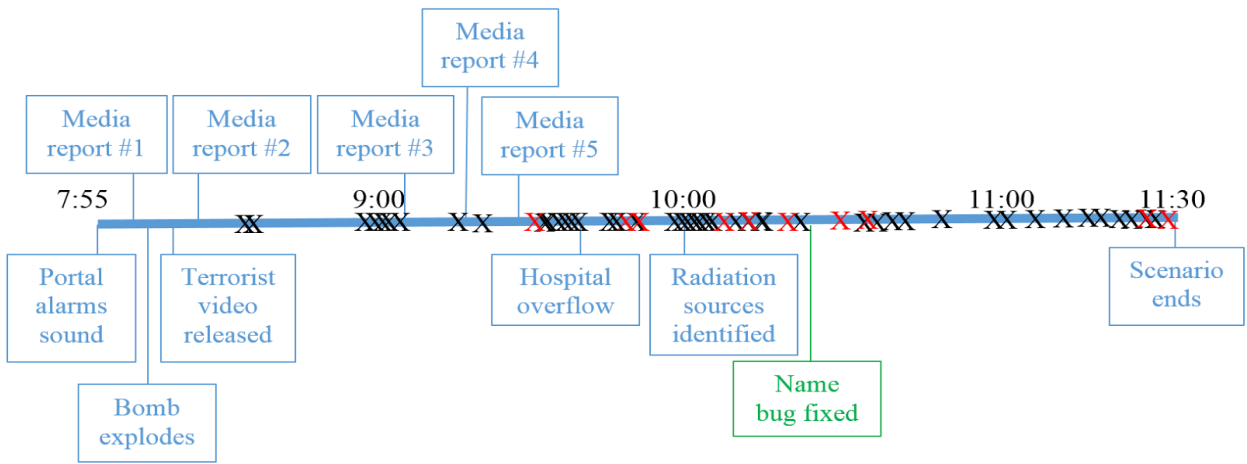

*Figure 4.* Exercise timeline. X's denote microblogs by PIO participants, red X's denote 'retweet' microblogs.

*6.2 Microblogs from PIOs*

    Six of the twenty-eight eligible PIOs actively engaged in microblog production, generating a total of 54 microblogs in the three hour exercise. An additional PIO reported discovery of the unannounced mobile device capability, and noted the multi-tasking advantage of this capability at the hot-wash. Several of the PIOs indicated technical problems using the interface. Two PIOs called for technical support and the problem was fixed relatively early, but others reported using incompatible browsers for the entire exercise, reducing or eliminating access and interaction with the social media messages.

The six actively blogging PIOs averaged one microblog every 23 minutes per active user, although our primary participant is responsible for 29 microblogs, with other users ranging from nine to one microblogs each. Figure 4 indicates the pattern of microblog production relative to the scenario events. The closely observed PIO developed a work practice of confirming a sent microblog by searching on its unique key-words. We also note substantial message editing to meet space constraints, while avoiding potential misunderstanding. This entailed abbreviation, replacement of process description with state information, imperative tense, removed politeness (e.g., "please"), preconditions (e.g., "if concerned"), and descriptors (e.g., "regional"). Complex microblogs may have been separated into two posts.

PIO Tweet content covered news (26%), general advice (50%) and medical advice (22%).

Example 7: News: Confirming an explosion at Hara Arena this morning. Several causalities (sic), injured. We will update as we get more information. #DaytonBombing
Example 8: General Advice: CDC recommends removing clothes and showering after exposure to radiation. Avoid things you have contacted while irradiated. #DaytonBombing
Example 9: Medical Advice: For those seeking medical attn:Decontamination units en route, please remain calm and be patient as they set up adjacent to area hospitals.

We note two different instances that solicit first responder personnel:

Example 10: We're in need of volunteer medical personnel. Report to the XXH main lobby w/ licensing info if you can help. #daytonbomb
Example 11: All off-duty #CityFire personnel: report to your home stations ASAP, await further instruction

Forty-four of the participant-generated microblogs were unique. The remaining ten were 'retweets' relaying both information content and pedigree. Figure 5 illustrates the resulting network of interactions between participants. Retweeted content included all of the above categories, including a single telephone number for obtaining further information.

Our primary PIO received no 'retweets' until after we announced we had fixed the naming bug via email. In the following twenty minutes, two different PIOs 'retweeted' his messages. This pattern of activity suggests both attention to the source and the influence of source on distribution. Furthermore, social media made the work of other physically distributed PIOs mutually visible and potentially enhanced coordination and consistency.

One way to determine the influence of our original social media stream on the actions of PIOs is to examine similarity in words and phrases produced. For example, our corpus of microblogs established the hashtag "#DaytonBombing" early in the simulation with both a high and a low visibility microblog. Within a half hour, our primary PIO had adopted this hashtag, including capitalization, and used it extensively. However, once the pre-constructed messages had finished propagating through the stream, other PIOs began to use "#daytonbomb" instead, a more efficient tag in a content-limited medium.

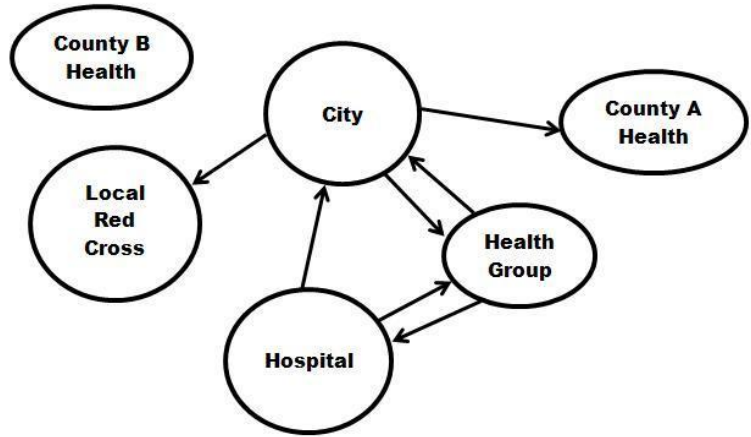

*Figure 5*. Network analysis of 'retweets'. Arrows indicate at least one message posted to the user account at the origin of the arrow, copied from the user account at the point of the arrow (e.g., the city 'retweeted' a message from the local Red Cross). The primary observed PIO is represented as "city".

We do see thematic cross talk between public blogs and PIO response. For example, the primary PIO attempted to locate a URL on a citizen blog with advice for describing the bombing episode to children. The PIO subsequently sent out Example 12, sharing the topic of children, but focusing on an entirely different issue:

Example 12: Parents: kids should follow same guidelines to control for any potential #radiation risk: stay indoors, shower, double-bag clothes

We also note a response to a public misconception regarding the lethality of the radioactive agent, initially identified in trending topics. At this point in the scenario, the radioactive agent has a name, but bloggers indicate that having a name for the agent is not helpful. The PIO responds with the following microblog:

Example 13: Dayton Fire Dept confirms that Tc-99m #radiation is NON-LETHAL, not considered dangerous. Still double-bag clothing, shower, stay indoors.

In so doing, the PIO addresses the lethality rumor, and also corrects the misstated name of the agent from the citizen's blog, from TC-98 to TC 99. Thus, the PIO shows engagement not by repeating blog content, but by correcting it.

Response to follow-up questionnaires was low and varied. While one PIO saw little benefit to engaging via social media, another volunteered to serve as a beta tester in future tool releases. Several acknowledged an effect on the procedures for interacting with other PIOs. One found the signal-to-noise ratio too high, while another found it to be realistic.

*6.3 Other Participant Interventions*

On two occasions, a scenario coordinator requested us to insert spontaneous microblogs, such as the following:

Example 14: @XXXCoHealth has no answers!!! #FreakedOutMom
Example 15: @XXXCoHealth has redeemed themselves. Their info hotline really helped!! #freakedOutMomNoMore

In Example 14, the scenario coordinator declared that a certain county had failed to provide a concerned citizen valuable information during a telephone exchange. The coordinator suggested that the researcher propagate Example 6 to the other PIOs to see how they reacted. We sent out the message and manually set the visibility to "medium" to approximate a relatively popular private Twitter user. We do not know if the chastised county saw Example 14. However, shortly thereafter (roughly thirteen minutes), the coordinator called the county again as a concerned citizen with more questions and this time approved of Darke County's handling of the situation. At the request of the scenario coordinator, we sent out Example 15 to praise the county performance.

We believe that these interventions, along with our participation on the design team itself, demonstrate the general enthusiasm of decision makers in the formal response community regarding the promise of social media. This enthusiasm persisted in comments during the hot-wash review, as well as the final recommendations to establish protocols and policies for the use of social media in disaster response.

## 7. Lessons Learned

While some participants questioned the realism of our signal-to-noise choice, realism in this regard was not our primary goal. We intentionally examined a social media flow rate and signal-to-noise ratio that appeared within the limits of human analysis capability. This allowed us to identify the opportunities for exploiting social media content in the absence of usability issues amenable to technology.

We observed the use of several features in the simple tool that we provided. Chief among these was the ability to respond to public concern and misconception directly. We also note an ability to copy a persisting written message and distribute it to the formal response team via e-mail. The ability to filter microblogs by trending topics made this process efficient--several microblogs could be copied and pasted at the same time. This replaces taking notes from telephone calls, limited not only by the ephemeral nature of audio messages, and the number of parallel answering machines, but also by the absence of explicit pedigree content available in users' social media profiles. However, work practice has not evolved to accommodate this capability, as evidenced by the lack of response via-email.

The availability of PIO microblogs and the ability to promote their distribution has the potential to benefit coordination across physically distributed PIOs. Social media renders their activities mutually intelligible, persistent and accessible in a unified medium. They can promote a consistent public response, for example pointing to CDC guidelines, or identifying a single regional point of contact for additional information.

While empirical papers generally focus on what was observed, we also note here the absence of behavior—what we did not see. Technical problems impacted broader PIO participation from the outset. Our web-based interface was built on the latest HTML5 technology and some web browsers types had partial support for this, causing several PIOs difficulty despite instructions. We do not view this problem as a permanent threat to deployment. This does however speak to the training requirements for the incorporation of comparable tools.

Participants will require experience with the computational capabilities in order to gain confidence. Our PIO first detected problematic microblogs primarily using the unfiltered stream and used the trending topics to select microblogs once a theme was recognized. We should expect persisting skepticism even with the use of computed trending topics, and a work practice that includes convergent evidence for any computational inference. This is the mark of a skilled operator [14].

Perhaps the most notable absence is related to the topics of engagement. This may be related to either the scope of PIO responsibilities, or the time span of the simulation. No one directly responded to the early misconceptions regarding the rejection of radiation patients using social media. We did not see attention to the estimation of medical need, although the microblog corpus included related content. We saw only modest attention to the distribution of medical resources, redirecting non-injured citizens away from hospitals as well as the local Air Force base, but not specific overloaded hospitals adjacent to the explosion. While these suggest the untapped potential of social media, they may also imply a future, fundamental change in work practice, in which other functions in addition to the PIO have access to public input to conduct their own discipline-related search.

## 8. Future Work

Demonstrating the potential function of social media in an operational setting justifies developing the capabilities of our more sophisticated social media analysis tool, Twitris (see Figure 6). For more than two years, members of our research team have been exploring the design of information tools for filtering social media to pull out potential signals (e.g., relevant or meaningful messages) from the noise [19] and for representing the relevant information in a useful format. Twitris features compensate for a realistic, high flow rate, and low signal to noise ratio that makes the unfiltered stream virtually unusable. Like the simple tool studied here, Twitris includes a trending topic capability, and the identification of related tweets. It also includes a mapping capability, and a search capability. Finally Twitris includes substantially expanded functions for identifying the role of the sender, to contribute to the assessment of trustworthiness. This can be important in the recognition and control of rumors and intentionally false postings by people who want to create panic and chaos, as happened in the Westgate Mall attacks [4].

The PIO in the present empirical work assumed responsibility for information filtering and distribution. *Twitris provides the ability for any emergency response function to monitor social media traffic*, selecting content according to interest. For example, tweets can be selected that pertain to the need for resources, a function outside the responsibilities of the PIO.

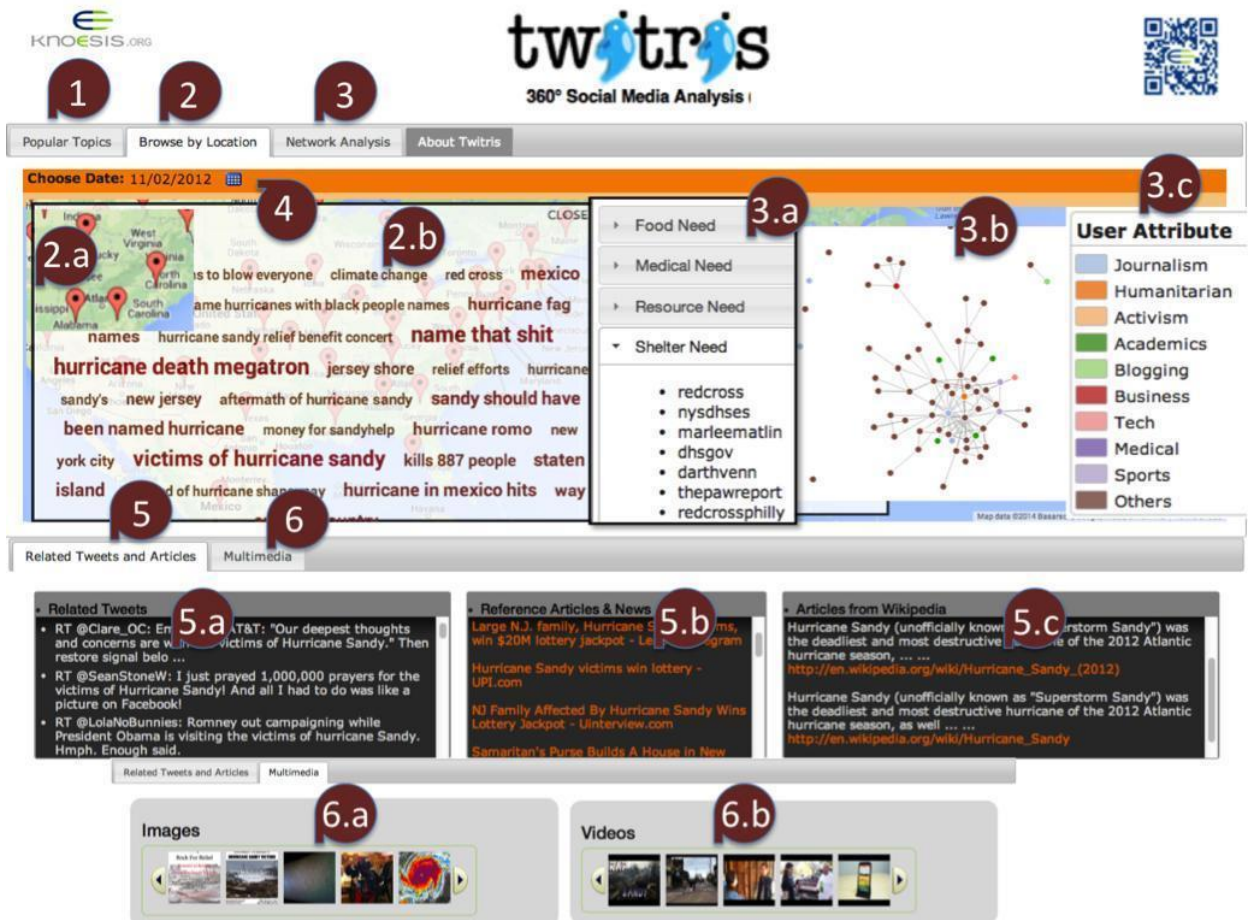

Figure 6: Twitris tool. Figure 1 shows the Twitris system where, (1) date-specific popular topical tags – also called social signals, clicking on which brings raw tweets to the topics [a Twitris search box invisible in the image also allows exploration over all the tweets regardless of topics based social signals]; (2) key social signals of discussions by regions – county, state, country: (2.a) to select location of interest—each pin shows a collection of social signals as in (2.b) emanating from that location; (3) topical user interaction networks in (3.b), and prioritized users for engagement in (3.a) with demographics, e.g., knowledge of user profession (3.c); (4) date of analysis; (5) display of tweets (5.a), recent news (5.b), and Wikipedia pages (5.c) related to a selected social signal; (6) event specific multimedia (images (6.a) and videos (6.b)).

**Conclusions**

Compared to the natural environment, training exercises offer control over the difference between critical variable (signal) and extraneous, uncontrolled confounds (noise). In addition, exercises make possible the recording of data and context relatively easily, as well as access to an a priori ground truth. However, this leaves open the question of whether our findings would differ under the pressures of a real emergency setting. While we have demonstrated a capability to exploit social media in emergency response, the full value requires information tools to filter the content of social media. When such information processing tools are available, they will change work practice.